\documentclass{ifacconf}

\usepackage{graphicx}      
\usepackage{natbib}        

\usepackage{float}
\usepackage{mathtools}
\usepackage{amsmath}

\usepackage{amssymb}
\usepackage{graphicx}
\usepackage{cases}

\newcommand{\scal}[2]{\langle #1,\, #2 \rangle}
\newenvironment{proof}[1][Proof]{\textbf{#1.} }{\ \rule{0.5em}{0.5em}}
\newtheorem{theorem}{Theorem}
\newtheorem{remark}{Remark}
\newtheorem{lemma}{Lemma}

\usepackage{xcolor}
\usepackage{amsmath,mathtools,amssymb}
\newcommand{\bi}{\begin{itemize}}

\newcommand{\ei}{\end{itemize}}
\newcommand{\bd}{\begin{description}}
\newcommand{\ed}{\end{description}}
\newcommand{\be}{\begin{enumerate}}
\newcommand{\ee}{\end{enumerate}}

\newcommand{\bqn}{\begin{eqnarray}}
\newcommand{\eqn}{\end{eqnarray}}
\newcommand{\eqnn}{\nonumber\end{eqnarray}}
\newcommand{\eqnl}[1]{\label{#1}\end{eqnarray}}

\newcommand{\ba}[1]{\begin{array}{#1}}
\newcommand{\ea}{\end{array}}

\newcommand{\R}{\mathbb{R}}

\newcommand{\cL}{\mathcal L}

\newcommand{\cX}{\mathcal X}

\def\be{{\bf e}}

\newtheorem{property}{Property}

\newtheorem{definition}{Definition}

\newtheorem{assumption*}{Assumption}

\def\cX{\mathcal{X}}

\newcommand{\ddt}{{\tfrac{{\rm d}}{{\rm d}t}}}

\usepackage{xcolor}
\newcommand{\beq}{\begin{equation}}
\newcommand{\eeq}{\end{equation}}

\newcommand{\bag}{\begin{aligned}}
\newcommand{\eag}{\end{aligned}}

 \newcommand{\E}{\mathrm{e}}

\begin{document}
\begin{frontmatter}

\title{On Hyperexponential Stabilization of Linear Infinite-Dimensional Systems} 

\author[First]{Moussa Labbadi} 
\author[First]{Christophe Roman}

\address[First]{Aix-Marseille University, LIS UMR CNRS 7020, 13013 Marseille, France (e-mail: \textsl{\{christophe.roman, moussa.labbadi\}@lis-lab.fr}). }

\begin{abstract} 
This paper study the hyperexponential stabilization for infinite-dimensional system on Hilbert space by a distributed time depending control law. The well-posedness of the closed loop for every time is obtained through the use of maximal monotone operator. The hyperexponential stability and ISS property of the closed loop is established using Lyapunov analysis and time scale transformation.
\end{abstract}   


\end{frontmatter}

\section{Introduction}

Non-autonomous dynamical systems have been studied since the 1950s for both finite-dimensional and infinite-dimensional systems, see \citep{caraballo2001decay,  carvalho2007characterization, carvalho2012attractors,song2024normal,bellin1953non}. In control theory, time-varying functions have been extensively utilized for various applications, including observer design \citep{efimov2022exact,wang2024exact}, hyperexponential stability analysis \citep{wang2024hyperexponential}, robustness analysis \citep{labbadi2024hyperexponential,labbadi2025hyperexponential}, and many other problems \citep{caraballo2001decay}.

The design of time-varying control laws for infinite-dimensional systems, particularly those governed by heat equations, was first proposed in \citep{carvalho2012attractors}. Similarly, time-varying control strategies have been applied to systems with time delays \citep{BellmanCooke1963, KolmanovskiiNosov1986, Gu2003, HaleLunel1993, Fridman2014}. However, the corresponding design procedures were initially non-constructive, and the well-posedness of the problem was not fully established.

The present paper extends this non-autonomous control method to infinite-dimensional control systems governed by abstract evolution equations, thereby broadening its applicability and providing a more robust framework for such systems.

Time-varying control algorithms, recognized for their accelerated stability and robustness, have recently been proposed for Linear Canonical Systems. Consider the diffusion process:
\[
u_t(t,x) = \nu u_{xx}(t,x) + U_1(t,x) + d(t), \quad x \in (a, b), \, t > 0
\]
with the control law:
\[
U_1(t,x) = -l \psi(t)^n u(t,x), \quad n\geq 1, \quad \psi(t) = 1+t,
\]
where $l>0$.
and the wave dynamics:
\[
u_{tt}(t,x) = \mu u_{xx}(t,x) + U_2(t,x) + d(t), \quad x \in (a, b), \, t > 0
\]
with the new control law for \( U_2(t) \):
\[
U_2(t,x) = -l_1 \psi(t)^{m_1} u(t,x) - l_2 \psi(t)^{m_2} u_t(t,x)
\]
where \( 1<m_1 < m_2 \) and \( l_1, l_2 \) are constants.

These systems are governed by time-varying controllers, under Dirichlet/Neumann boundary conditions. The scalar state \( u(t,x) \) is a function of the spatial variable \( x \in [0, 1] \) and time \( t \geq 0 \). The inputs \( U_1 \) and \( U_2 \) are distributed control inputs, while \( d_1 \) and \( d_2 \) represent matched external disturbances. These in-domain stabilization procedure holds also interest in the context of boundary control because these time-varying closed loop system can be used as a target system in the context of backstepping \cite{vazquez2026backstepping}.

It has been proven that for linear systems finite dimensional system~\citep{wang2024exact,wang2024hyperexponential,labbadi2024hyperexponential}, the rapid convergence and reject of perturbations are obtained. Up to our knowledge, sufficient conditions ensuring the same result in the infinite-dimensional case do not exist. Addressing this gap is the objective of the present paper.

The main contributions of this paper are summarized as follows:
\begin{enumerate}
\item We establish the existence and uniqueness of mild solutions for a broad class of infinite-dimensional systems by leveraging the Duhamel formula and the Banach–Picard fixed-point theorem.
\item We design linear time-varying feedback controllers, constructed via Lyapunov functions, that ensure input-to-state stability (ISS) together with hyperexponential convergence.
\item We integrate time-scale techniques with Lyapunov-based arguments to derive hyperexponential stability conditions for time-varying controllers in infinite-dimensional settings.
\item We demonstrate the effectiveness of the proposed framework through a detailed study of a one-dimensional heat equation with an exponential-kernel memory term.
\end{enumerate}
The rest of the paper is organized as follows. Section~\ref{sec:preliminaries} reviews preliminaries on contraction semigroups and maximal monotone operators. Section~\ref{sec:main_results} establishes the well-posedness of the closed-loop system, presents the control design, and analyzes the resulting hyperexponential stability. Section~\ref{sec:Proof_main_results} contains the proofs of the main results. Section~\ref{application} illustrates the proposed approach through an application to the heat equation with an exponential memory term and provides numerical results. Finally, Section~\ref{sec:conclusion} concludes the paper.

\subsection*{Notation}
\begin{itemize}
    \item \( \mathbb{R}_+ = \{x \in \mathbb{R} : x \geq 0\} \) and \( \mathbb{R}^*_+ = \{x \in \mathbb{R} : x > 0\} \), where \( \mathbb{R} \) represents the set of real numbers.
    
    \item The absolute value in \( \mathbb{R} \) is denoted by \( | \cdot | \), and \( \| \cdot \| \) represents the Euclidean norm on \( \mathbb{R}^n \).
    
    \item For a (Lebesgue) measurable function \( d : \mathbb{R}_+ \to \mathbb{R}^m \), the norm \( \|d\|_\infty = \text{ess sup}_{t \ge 0} \|d(t)\| \) is defined. The set of functions \( d \) satisfying \( \|d\|_\infty < +\infty \) is denoted as \( L^\infty_m \).
    
    
    \item A continuous function \( \alpha : \mathbb{R}_+ \to \mathbb{R}_+ \) belongs to the class \( \mathcal{K} \) if \( \alpha(0) = 0 \) and it is strictly increasing. A function \( \alpha : \mathbb{R}_+ \to \mathbb{R}_+ \) is in the class \( \mathcal{K}_\infty \) if \( \alpha \in \mathcal{K} \) and \( \alpha \) increases to infinity. A continuous function \( \beta : \mathbb{R}_+ \times \mathbb{R}_+ \to \mathbb{R}_+ \) is in the class \( \mathcal{KL} \) if \( \beta(\cdot, t) \in \mathcal{K} \) for each fixed \( t \in \mathbb{R}_+ \) and \( \beta(s, \cdot) \) decreases to zero for each fixed \( s >0 \).
    
\end{itemize}
\section{Preliminaries}\label{sec:preliminaries}
In this section, we review contraction semigroups and maximal monotone operators, and we recall an auxiliary lemma used in our analysis.
\subsection{Semi-groups of contraction and maximal monotone operator}
The context of this paper is partial differential equations. To study well-posedness, we require  definitions and results from abstract evolution problems in functional spaces, typically Banach spaces and, in particular, Hilbert spaces. Indeed, partial differential equations can be viewed as differential equations in functional spaces.

As this paper consider nonlinear operator it seams necessary to present some of the notation, the operator is identified with it Graph, which is denoted by $\mathcal{G}(\cdot)$, $Gx=\{y:(x,y)\in \mathcal{G}(G)\}$. The domain of the operator is $\mathcal{D}(G)=\{ x: Gx\neq \varnothing \}$. The range of the operator is $\mathcal{R}(G)=\bigcup_{x\in\mathcal{D}(G))} Gx$. The inverse of an operator is defined as $(y,x)\in \mathcal{G}(G^{-1}) \Leftrightarrow (x,y)\in \mathcal{G}(A)$. A linear operator $G$ in the Hilbert space $\mathtt{H}$ is called monotone if and only if, $\forall x\in \mathcal{D}(G)$,
\begin{equation}
 \scal{Gx}{x}\geqslant 0.
\end{equation}

\begin{property}
Consider $G$ a operator in $\mathtt{H}$ the following sentences are equivalent
\begin{itemize}
\item $G$ is maximal monotone
\item $G$ is monotone and $\mathcal{R}(I+G)=\mathtt{H}$
\item $\forall \lambda >0$, $(I+\lambda G)^{-1}$ is a contraction defined on all $\mathtt{H}$
\end{itemize}
\end{property}
This can be found in \citep{brezis2010functional}. 
\begin{definition}[according to \citep{komura1967nonlinear}]\label{def_semigroups}
    Let $\mathtt{H}$ be a Hilbert space, a one parameter nonlinear semigroups of contraction $\{S(t),t\geq 0\}$ is defined by
    \begin{itemize}
        \item $\forall t \geq 0$, $S(t)$ is a continuous nonlinear operator defined on $\mathtt{H}$ into $\mathtt{H}$.
        \item $\forall t \geq 0, \forall x\in\mathtt{H}$, $S(t)x$ is strongly continuous in $t$.
        \item $\forall t,s\geq 0$, $S(t+s)=S(t)S(s)$, and $S(0)=I$ where $I$ stands for identity mapping.
        \item $\Vert{S(t)x-S(t)y}\Vert\leq \Vert{x-y}\Vert$ for every $x,y\in\mathtt{H}$.
    \end{itemize}
\end{definition}
 Moreover, the reader will find in \citep{komura1967nonlinear} an equivalence between the concept of nonlinear contraction semigroups as defined before and maximal monotone operators. In brief, if $A$ is maximal monotone in $\mathtt{H}$, then it exist a contraction semigroups $S$ such that for each $x\in H$ then $S(t)x\in H$ existence of weak solution, moreover for each $x\in \mathcal{D}(A)$ then $S(t)x\in \mathcal{D}(A)$ existence of strong solution.  These are classical and well-established results within the mathematical analysis community. $G$ is maximal montone is equivalent to $-G$ is m-dissipative and to $G$ is m-accretive.
\subsection{Auxiliary properties}
In this paper, the following lemma will be used.
\begin{lemma}\label{lem:Update_3} \citep{labbadi2024hyperexponential}
For all \( \tau \geq 0 \), and $a,\alpha>0$ such that \( \alpha a > 1 \), the following holds:
\[
\int_{0}^{\tau}\mathrm{e}^{s-\tau}\frac{ds}{(a s + 1)^\alpha} \leq \frac{r_{a,\alpha}}{(a \tau + 1)^\alpha},
\]
where
\[
\begin{aligned}
r_{a,\alpha} &= (a \alpha)^{\alpha} \int_{0}^{\frac{a\alpha - 1}{a}} \mathrm{e}^{s - \frac{a\alpha - 1}{a}} \frac{ds}{(a s + 1)^{\alpha}}  \\
&\qquad + (a \alpha + 1) + \left(\frac{a \alpha + 1}{a \alpha}\right)^{\alpha} \left(1 - \mathrm{e}^{-\frac{1}{a}}\right).
\end{aligned}
\]
\end{lemma}
\section{Main results}\label{sec:main_results}
\label{sec:problem}
Considering a given Hilbert space \(\cX\) equipped with a scalar product \(\langle \cdot, \cdot \rangle_\cX\) and the induced norm \(\|\cdot\|_\cX\). We focus on the abstract control framework for the following non-autonomous problems of the form:
\begin{equation}\label{eq:abstractform}
\left\{
\begin{aligned}
\ddt X(t)+A X(t) &= f(t)X(t)+d(t), \\
X(0) & = X_0,
\end{aligned}
\right.
\end{equation}
\begin{theorem}\label{theorem_well_posed}
    Considering the abstract evolution problem \eqref{eq:abstractform}. If $A$ is linear maximal monotone, $f\in L^\infty_{loc}(\mathbb{R}^+;\cX)$ and $d\in L^1_{loc}(\mathbb{R}^+;\cX)$, then for every $X_0\in\overline{\mathcal{D}(A)}=\cX$ it holds for all $T>0$
    \begin{align}
        X\in C^0([0,T],\cX).
    \end{align}
    In the more regular case, if $f$ is Lipschitz continous and $d$ is Lipschitz continous, then for every $X_0\in{\mathcal{D}(A)}$ it holds for all $T>0$
    \begin{align}
        X\in C^0([0,T];{\mathcal{D}(A)})\cap C^1([0,T];\cX).
    \end{align}
\end{theorem}
The well-posedness of \eqref{eq:abstractform} holds when $f(t)$ is Lipschitz continuous (with $d=0$) is a well known result \cite[Theorem 1.2 and Theorem 1.5, Theorem 1.6 In Chapter 6]{pazy1983semigroups} different hypothesis of $f$ yields different result. One can integrate $d$ into $f$ but then the regularity of $d$ is constraint. In order to make relax hypotheses on the regularity of $d$ we perform classics computation in the case of the mild solution, the strong solution result is directly \cite[Theorem 1.6 In Chapter 6]{pazy1983semigroups}. We still remark that there exist various result in the direction of nonautonoumous abstract evolution problem the work of Pavel \citep{pavel2006nonlinear}, Kobayasi et al., Evans \citep{evans1977nonlinear}, and Crandall and Pazy, as well as Kato's work on perturbation theory \citep{kato2013perturbation}. 

Our main goal now that solution are indeed well-defined is to design a time-varying state feedback control $f$ that renders the system input-to-state stable (ISS) \citep{sontag2007input}, with a hyperexponential rate of convergence (faster than any exponential rate) under the constraint. 

\subsection{Uniform Hyperexponential Stability}
Standard stability concepts for nonautonomous nonlinear systems described by partial differential equations (PDEs), such as asymptotic stability and input-to-state stability, are well-established in the literature. In this paper, we introduce the less conventional concept of uniform hyperexponential stability, relevant for systems in Hilbert spaces.
The following definition introduces uniform hyperexponential stability for the system:
\begin{defn}
The system \eqref{eq:abstractform} is uniformly hyperexponentially stable if there exist constants \( \kappa_0 \in \mathbb{R}_+ \), functions \( \rho, \kappa \in \mathcal{K}_\infty \), and \( \beta \in \mathcal{KL} \) such that:
\begin{equation}\label{eq:s_hyp}
\bag
\|\varphi(t, t_0, X_0, d)\| &\leq \E ^{\left( -(\kappa(t - t_0) + \kappa_0)(t - t_0) \right)} \rho(\|X_0\|) \\
& \qquad+ \beta(\|d\|_\infty, t - t_0),
\eag
\end{equation}
for all \( t \geq t_0 \), \( X_0 \in \mathcal{X} \), and \( d \in D \).
\end{defn}

Here:
 \( \|\varphi(t, t_0, X_0, d)\| \) is the norm of the solution at time \( t \). The exponential term \( \E^{\left( -(\kappa(t - t_0) + \kappa_0)(t - t_0) \right) }\) captures the uniform hyperexponential rate of convergence. \( \rho(\|X_0\|) \) depends on the initial state. \( \beta(\|d\|_\infty, t - t_0) \) accounts for the disturbance.

This stability guarantees rapid convergence and robustness to disturbances, with a uniform rate independent of initial conditions and disturbances within specified sets.
\subsection{Feedback control design} 
It is requested to design a robust controller that can handle matched perturbations while achieving faster convergence than any exponential rates.
We propose the following time-varying control law:
\begin{equation}
    u(t) = -K\psi(t)^n X(t), \label{eq:u}
\end{equation}
where   $\psi(t) = 1+t $ is a time function, $K>0$ and $n\geq 1$. 
\begin{remark}
Any continuous, strictly increasing function \( \psi : \mathbb{R}^*_+ \to \mathbb{R}^*_+ \) with \( \psi(0) > 0 \) and an unbounded integral can be utilized in equation \eqref{eq:u}. For instance, \( \psi(t) = a\E^{\alpha t}) \) or \( \psi(t) = bt^t \), where \( a > 0 \), \( b > 0 \), and \( \alpha > 0 \). By applying a change of time \citep{wang2024exact,wang2024hyperexponential,efimov2022exact}, all these functions can be transformed into the one used for the proof in this work.
\end{remark}
In this case, the closed-loop system is
\begin{equation}\label{eq:abstractform_closed-loop}
\left\{
\begin{aligned}
\ddt X(t)+A X(t) &= -K\psi(t)^n X(t)+d(t), \\
X(0) & = X_0,
\end{aligned}
\right.
\end{equation}
 We now state the first result on convergence and robustness.
\begin{theorem}\label{thm:convergence}
Assume that \(A\) is an \(m\)-accretive operator on \( \mathcal{X} \times \mathcal{X} \) and that \( K > \tfrac{1}{2} \). Then, for every initial condition \( X_0 \in \overline{D(A)} \), the closed-loop system~\eqref{eq:abstractform_closed-loop} is uniformly ISS and exhibits hyperexponential stability.
\end{theorem}
\subsection{Partial Stability}

Consider the closed-loop system
\begin{equation}\label{eq:abstractform_closed-loop_B}
\left\{
\begin{aligned}
\dot X(t)+A X(t) &= -K\psi(t)^n B X(t) + d(t),\\
X(0) &= X_0.
\end{aligned}
\right.
\end{equation}

\begin{theorem}[Stability and ISS]\label{thm:stability}
Assume $A$ is maximal monotone, $B$ is bounded and coercive, and $\psi$ is continuous with $\psi(0)>0$.  
If $K\beta>\tfrac12$ so that $A+K\psi(t)B$ is $m$-accretive for all $t\ge0$, then for every $d\in L^\infty(\R_+;\mathcal{X})$ and $X_0\in\overline{D(A)}$, the mild solution satisfies
\begin{align}\label{eq:ISS_bound}
\|X(t)\|^2
\le
\E^{-\eta t(t+2)/2}\,\|X_0\|^2
+\frac{C}{\psi(t)^2}\,\|d\|_\infty^2,
\qquad t\ge0,
\end{align}
with $\eta:=2K\beta-1>0$ and $C>0$ independent of $t$.

Thus the homogeneous dynamics decay hyperexponentially, and the closed-loop system~\eqref{eq:abstractform_closed-loop_B} is ISS with respect to bounded disturbances.
\end{theorem}

\section{Proof of the main results}\label{sec:Proof_main_results}
In the following one can find the proof of Theorem~\ref{theorem_well_posed}.

\begin{proof}
First as $A$ is maximal monotone this implies that $-A$ is the generator of a strongly continuous semigroups of contraction \citep{komura1967nonlinear} denoted $S$. Second using the Duhamel formulae it holds
\begin{align}
    X(t)=&S(t)X_0+\int_0^t S(t-s)(f(s)X(s)\notag\\&+d(s))ds.
\end{align}

Let $t_0 \in [0,T)$ and $X(t_0)\in \cX$.  
We consider the Banach space $C([t_0,T];\cX)$ equipped with the norm
\[
\|v\|_{C([t_0,T];\cX)}
   := \sup_{t\in[t_0,T]}\|v(t)\|_{\cX}.
\]
Assume that $(S(t))_{t\ge 0}$ is a $C_0$-semigroup on $\cX$ satisfying 
$\|S(t)\|\leq 0$, that  
$f(\cdot):[0,T]\to \cL(\cX)$ is locally bounded, and that 
$d(\cdot)\in C([t_0,T];\cX)$.

For $v\in C([t_0,T];\cX)$ define the operator $F$ such that $\forall t \in[t_0,T]$
\begin{align}
(Fv)(t)
    =& S(t-t_0) X(t_0)\notag\\&
       + \int_{t_0}^{t} S(t-s)\bigl( f(s)v(s)+d(s)\bigr)\,ds,
\end{align}
A function $u\in C([t_0,T];\cX)$ is a mild solution of the evolution equation
\[
\dot u(t)=Au(t)+f(t)u(t)+d(t),
\qquad u(t_0)=X(t_0),
\]
if and only if $u$ is a fixed point of $F$.

\medskip
Let $v,w\in C([t_0,T];\cX)$ for some $T\in(t_0,T]$.  
Then for every $t\in[t_0,T]$,
\begin{align}
\|(Fv)(t)-(Fw)(t)\|
&\le \int_{t_0}^{t}
   \|S(t-s)\|\,\|f(s)\|\notag\\& \times \|v(s)-w(s)\|\,ds .
\end{align}
Taking the supremum over $s\in[t_0,T]$ yields
\begin{align}
\|Fv-Fw\|_{C([t_0,T];\cX)}
   &\leq 
     \sup_{t\in[t_0;T]}\int_{t_0}^{t} \|f(s)\|\,ds\notag\\ & \times
     \|v-w\|_{C([t_0,T];\cX)} .
\end{align}
Since $f(\cdot)$ is locally bounded it holds $\int_{t_0}^T \|f(s)\|\,ds\leq M (T-t_0)$.
Thus choosing $t_0+\frac{2}{M}<T_1<\frac{1}{M}+t_0$ such that $\int_{t_0}^{T_1} \|f(s)\|\,ds<1$, shows that $F$ is a contraction on $C([t_0,T_1];\cX)$. By the Banach-Picard fixed-point theorem, $F$ has a unique fixed point on this interval. By iterating this construction on successive short intervals $[t_0,T_1], [T_1,T_2],\dots$of minimum length $\frac{2}{M}$, one obtains a unique global mild solution $u\in C([t_0,T];\cX)$.

As previously noted, the existence and uniqueness of a strong solution follow directly from the application of \cite[Theorem 1.6, Chapter 6]{pazy1983semigroups}. But it should be possible to narrow the Lipschitz hypothesis.
\end{proof}

The proof of Theorem~\ref{thm:convergence} is presented as follows:

\begin{proof}
The dynamics of the state \( X(t) \in \mathcal{H} \) are governed by the differential equation:
\[
\frac{d}{dt} X(t) = -A X(t) - K\psi(t) X(t) + d(t),
\]
Define the Lyapunov function:
\[
V(t) = \langle X, X\rangle.
\]
those time derivative  is:
\begin{gather*}
\frac{d}{dt} V(t) = 2 \left\langle \frac{dX}{dt}, X\right\rangle \\
 = 2 \left\langle -A X(t) - K\psi(t) X(t) + d(t), X(t) \right\rangle\\
 = -2 \left\langle A X(t), X(t) \right\rangle - 2K\psi(t) V(t) + 2 \left\langle d(t), X(t) \right\rangle.
\end{gather*}
Using Young's inequality 
\[
2 \left\langle d(t), X(t) \right\rangle \leq\psi^{-1} \|d(t)\|^2 + \psi \|X(t)\|^2.
\]
Substituting this into the expression for \( \frac{d}{dt} V(t) \):
\[
\frac{d}{dt} V(t) \leq - \psi(t)(2K-1) V(t) + \psi^{-1}\|d(t)\|^2.
\]
If $K > 0.5$, then we define $\eta = 2K-1>0$. 
To refine the analysis, we introduce a new time variable defined by the transformation:
\[
d\tau = \eta\psi(t) \, dt, \quad \text{which leads to} \quad \tau = \varphi(t) = \eta \frac{t}{2} \left( t + 2 \right).
\]
The inverse relation is given by:
\[
t = \varphi^{-1}(\tau) = \sqrt{2\eta^{-1}\tau + 1} - 1, \quad \text{with} \quad \psi(\tau) = \sqrt{2\eta^{-1}\tau + 1}.
\]

The Lyapunov function in the transformed time domain is expressed as:
\[
V(\tau) = V\left[ \varphi^{-1}(\tau) \right],
\]
and its derivative satisfies:
\beq\label{eq:Ax1}
\begin{aligned}
\frac{dV(\tau)}{d\tau} &\leq \frac{dV(t)}{dt} \bigg|_{t = \varphi^{-1}(\tau)} \frac{d\varphi^{-1}(\tau)}{d\tau} \\
&\leq - V(\tau) + \frac{d^2(\tau)}{4\eta(2 \eta^{-1} \tau + 1)}.
\end{aligned}
\eeq

By solving this inequality, we derive the following integral representation for \( V(\tau) \):
\beq\label{eq:x1_tau}
\begin{aligned}
V(\tau) \leq \E^{-\tau} V(0) + \int_0^\tau \E^{s-\tau} \frac{d^2(\tau)}{4\eta \psi(s)^2} \, ds,
\end{aligned}
\eeq
with an upper bound given by:
\beq\label{eq:x1_tau_Up}
\begin{aligned}
V(\tau) \leq \E^{-\tau} V(0) + \int_0^\tau \E^{s-\tau} \frac{\|d\|_{\infty}^2}{4\eta \psi(s)^2} \, ds.
\end{aligned}
\eeq

Using Lemma \ref{lem:Update_3}, the upper bound of \( V(\tau) \) satisfies:
\[
V(\tau) \leq \E^{-\tau} V(0) + \frac{4}{\eta} \frac{r_{\frac{2}{\eta},1} \|d\|_\infty^2}{\frac{2\tau}{\eta} + 1}.
\]

Reverting to the original time variable \( t \), the corresponding upper bound for \( V(t) \) is:
\[
V(t) \leq \E^{-\eta \frac{t}{2} \left( t + 2 \right)} |V(0)| + \frac{4}{\eta} \frac{r_{\frac{2}{\eta},1} \|d\|_\infty^2}{\psi(t)^2}.
\]

The second term, \(\frac{4}{\eta} \frac{r_{\frac{2}{K},1} \|d\|_\infty^2}{\psi(t)^2}\), vanishes as \( t \to \infty \). Thus, the Lyapunov function \( V \) is uniformly bounded and demonstrates hyperexponential convergence to zero. This rate of convergence surpasses traditional exponential decay, offering enhanced robustness and rapid stabilization. 
Returning on the state \( X \), we have:
\[
\|X(t)\|^2 \leq \E^{-\eta \frac{t}{2} \left( t + 2 \right)} X(0)^2 + \frac{4}{\eta} \frac{r_{\frac{2}{\eta},1} \|d\|_\infty^2}{\psi(t)^2}.
\]
This completes the proof. 
\end{proof}

In the following, we provide the proof of Theorem~\ref{thm:stability}. 

\begin{proof}
We employ the Lyapunov function $V(t):=\langle X, X\rangle$. Differentiating along trajectories (justified for mild/strong solutions by standard approximation arguments) gives
\begin{gather*}
\dot V(t) = 2\langle \dot X(t),X(t)\rangle \\ 
= -2\langle A X(t),X(t)\rangle -2K\psi(t)\langle B X(t),X(t)\rangle + 2\langle d(t),X(t)\rangle.
\end{gather*}
Monotonicity of $A$ implies $\langle A X,X\rangle\ge0$. Using the coercivity of $B$,
\[
-2K\psi(t)\langle B X,X\rangle \le -2K\beta\,\psi(t)\|X\|^2.
\]
Apply Young's inequality with weight $\psi(t)>0$ to the disturbance term:
\[
2\langle d,X\rangle \le \psi(t)\|X\|^2 + \psi(t)^{-1}\|d\|^2.
\]
Combining the three estimates yields
\[
\dot V(t) \le -\psi(t)(2K\beta-1)V(t) + \psi(t)^{-1}\|d(t)\|^2.
\]
Set $\eta:=2K\beta-1>0$. Then
\[
\dot V(t) \le -\eta\,\psi(t)\,V(t) + \psi(t)^{-1}\|d(t)\|^2. 
\]

Perform the time reparametrization
\[
d\tau = \eta\psi(t)\,dt, \qquad \tau=\varphi(t)=\eta\int_0^t\psi(s)\,ds.
\]
For the model choice $\psi(t)=1+t$ we have the explicit relation
\[
\tau=\tfrac{\eta}{2}t(t+2),\qquad t=\sqrt{2\eta^{-1}\tau+1}-1,
\]
and $\psi(t)^2=2\eta^{-1}\tau+1$ when expressed in $\tau$.

Define $V(\tau):=V(\varphi^{-1}(\tau))$. Using (1) and the chain rule,
\[
\frac{dV}{d\tau}(\tau)
\le -V(\tau) + \frac{\|d(\varphi^{-1}(\tau))\|^2}{\eta\,\psi(\varphi^{-1}(\tau))^2}.
\]
Let $h(\tau):=\frac{\|d(\varphi^{-1}(\tau))\|^2}{\eta\,\psi(\varphi^{-1}(\tau))^2}$. By variation of constants,
\[
V(\tau)\le \E^{-\tau}V(0) + \int_0^\tau \E^{-(\tau-s)}h(s)\,ds.
\]
Using the uniform bound $\|d(\cdot)\|\le\|d\|_\infty$ and $\psi(\varphi^{-1}(s))^2=2\eta^{-1}s+1$ we obtain
\[
V(\tau)\le \E^{-\tau}V(0) + \frac{\|d\|_\infty^2}{\eta}\int_0^\tau \frac{\E^{-(\tau-s)}}{2\eta^{-1}s+1}\,ds.
\]
From Lemma~\ref{lem:Update_3}, multiplying both sides by $(2\eta^{-1}\tau+1)$ and rearranging shows that the integral term is bounded by
\[
\frac{\|d\|_\infty^2}{\eta}\cdot \frac{C(\eta)}{2\eta^{-1}\tau+1},
\]
where
\[
C(\eta):=\sup_{\tau\ge0}\Big\{(2\eta^{-1}\tau+1)\int_0^\tau \frac{\E^{-(\tau-s)}}{2\eta^{-1}s+1}\,ds\Big\},
\]
and $C(\eta)<\infty$ (the integrand is bounded and the exponential kernel yields uniform integrability; a simple dominated convergence argument shows the supremum is finite). Returning to original time $t$ using $2\eta^{-1}\tau+1=\psi(t)^2$ yields \eqref{eq:ISS_bound}.

Thus the homogeneous term decays like $\E^{-\eta\tfrac{t}{2}(t+2)}$ (hyperexponential) and the disturbance term is $O(\psi(t)^{-2})$, which vanishes as $t\to\infty$. This proves hyperexponential stability of the nominal closed-loop and ISS with respect to bounded matched disturbances. 
\end{proof}

\section{Application: Heat Equation with Memory Term}\label{application}

To illustrate the method, we apply the results to a heat equation inspired from heat equation with memory term.

\subsection{Problem Statement and Reformulation}
Consider the system :
\begin{subnumcases}{\label{sys_heat}}
v_{t}(t,x) = v_{xx}(t,x) - w(t,x) + U_1(v(t,\cdot),x) + d(t),  \\
v(t,0) = v(t,1) = 0, \\
w_t(t,x) = -\beta w(t,x) + \eta v(t,x)+ \varepsilon U_2(v(t,\cdot),x),\\
v(0,\cdot) = v_0, \quad w(0,\cdot) = w_0.
\end{subnumcases}
Using the variation of constants formula, without lose of generality we have when $\varepsilon=0$
\begin{align}
    w(t,x) = \E^{-\beta t} w_0(x) + \eta \int_0^t \E^{-\beta(t-s)} v(s,x) \, ds.
\end{align}

The equation \eqref{sys_heat} can then be rewritten when $\varepsilon=0$ as
\begin{align}\label{memory}
v_t(t,x) = v_{xx}(t,x) - \E^{-\beta t} w_0(x) \notag \\ 
- \eta \int_0^t \E^{-\beta(t-s)} v(s,x) \, ds 
+ U(v(t,\cdot),x) + d(t).
\end{align}
with
\begin{align}
    U(v(t,\cdot),x)=&\int_0^t \E^{-\beta(t-s)}\varepsilon U_2(v(s,\cdot),x)ds+U_1(v(t,\cdot),x)\notag
\end{align}
The integral term represents a memory contribution with an exponential kernel. The initial term $w_0$ can be set to zero in order to recover the classical memory formulation commonly used in the literature. Indeed, equations~\eqref{memory} with perturbations play a fundamental role in modeling numerous physical phenomena, including viscoelasticity and heat conduction. Their origins can be traced back to the pioneering works of \cite{Maxwell1867},  \cite{Boltzmann1874,Boltzmann1878}, and  \cite{Volterra1912,Volterra1913}. In the context of elastic materials, Boltzmann and Volterra expressed the stress tensor in terms of both the instantaneous strain and its historical values, thereby capturing the material’s memory effects. Equations incorporating memory terms have been extensively studied; see, for example, \citep{Amendola2012,Cattaneo1958,ChavesSilva2017,Christensen1982,Coleman1967,Dafermos1970,Fabrizio2010,Fu2009,Gurtin1968,Lu2017,Pandolfi2018}. Notably,  \cite{Gurtin1968} analyzed the general memory effect in heat conduction, showing that temperature waves propagating along the heat-flux direction travel faster than those moving oppositely. Similarly,  \cite{Dafermos1970} investigated the asymptotic behavior of linear viscoelastic systems at large times, introducing an auxiliary variable to account for the history of the states.

 Defining the state
\begin{align}
X(t) = \begin{bmatrix} v(t,\cdot) \\ w(t,\cdot) \end{bmatrix},
\end{align}

and the linear time-varying input operator
\begin{align}\label{def_U1_reform}
U(v(t,\cdot)) = -K \psi(t) v(t,\cdot),
\end{align}

with the coercive operator
\begin{align}\label{def_B_reform}
B_\epsilon = \begin{bmatrix} 1 & 0 \\ 0 & \epsilon \end{bmatrix}, \quad \epsilon > 0,
\end{align}

and the operator \(A\) defined by
\begin{gather*}
Az = \begin{bmatrix} -z_1'' + z_2 \\ \beta z_2 - \eta z_1 \end{bmatrix}, \\ 
\mathcal{D}(A) = \{ z \in H^2 \times L^2 \ | \ z_1(0) = z_1(1) = 0 \},
\end{gather*}

the abstract evolution equation can now be written in the standard linear feedback form
\begin{align}\label{abstract_form}
\dot X + A X + K \, \psi(t) \, B_\epsilon X + d(t) = 0,
\end{align}
where 
\begin{align}
\psi(t) = 1 + t.
\end{align}

The operator \(A\) is monotone in the Hilbert space \(L^2 \times L^2\) with the inner product
\begin{align}
\langle z, q \rangle = \int_0^1 \eta z_1 q_1 + z_2 q_2 \, dx,
\end{align}
and satisfies the range condition, ensuring it is maximal monotone.

\begin{remark}
 The operator $B_\epsilon$ is coercive in $L^2 \times L^2$, since
\[
\langle B_\epsilon z, z \rangle = \int_0^1 \eta z_1^2 + \epsilon z_2^2 \, dx \ge c \|z\|^2, \quad c = \min(1,\epsilon) > 0.
\]\end{remark}

The abstract system \eqref{abstract_form} is equivalent to the original heat equation with memory term, with the linear time-varying feedback \(U(v) = -K \psi(t) v\) enforcing rapid stabilization despite the memory term.
\subsection{Results}
\begin{figure}[htb!]
    \centering
    \includegraphics[width=.9\linewidth]{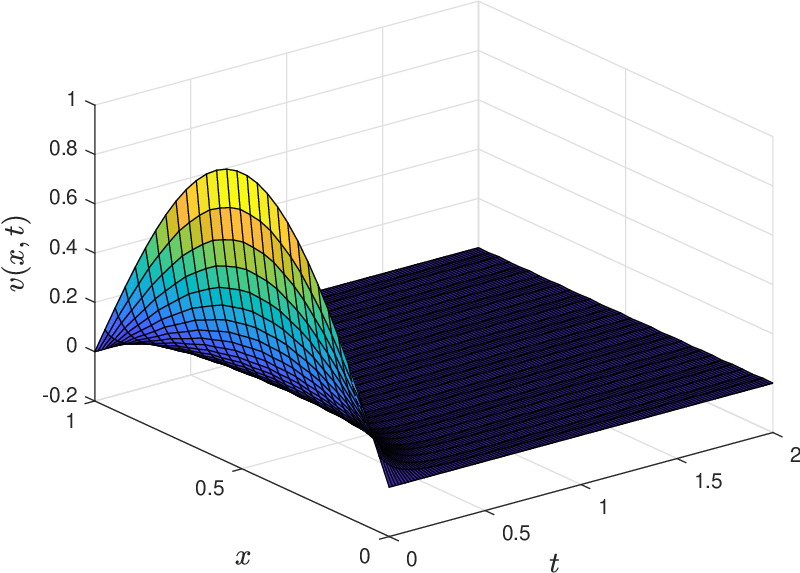}
    \caption{Distributed state $v(t,\cdot)$.}
    \label{fig_1}
\end{figure}
\begin{figure}[htb!]
    \centering
    \includegraphics[width=.9\linewidth]{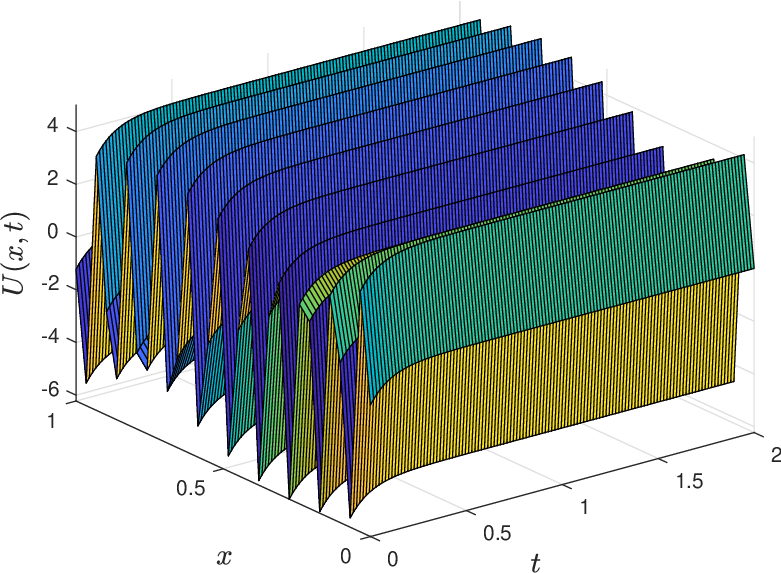}
    \caption{Control $U(t)$.}
    \label{fig_2}
\end{figure}
\begin{figure}[htb!]
    \centering
    \includegraphics[width=.9\linewidth]{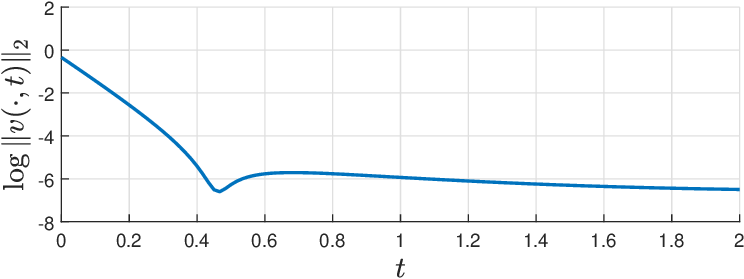}
    \caption{Objective $\log(\|v(t,\cdot)\|)$.}
    \label{fig_3}
\end{figure}
\begin{figure}[htb!]
    \centering
    \includegraphics[width=0.9\linewidth]{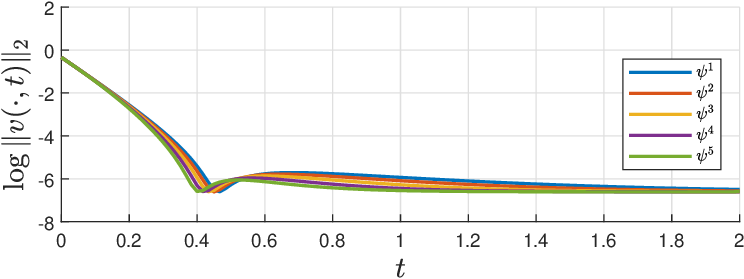}
    \caption{Evolution of the objective function $\log\!\left(\|v(t,\cdot)\|_{2}\right)$ for $\psi^n$ with $n \in \{1,2,3,4,5\}$.}
    \label{fig:logL2_psi_n}
\end{figure}

The simulations illustrate the effectiveness of the linear time-varying feedback 
in stabilizing the heat equation with memory. 
Figures~\ref{fig_1} and~\ref{fig_2} show the evolution of the distributed state 
$v(t,\cdot)$ and the corresponding control $U(v(t,\cdot),t)$, respectively. 
The decay of the objective function 
$\log\!\left(\|v(t,\cdot)\|_{2}\right)$ for different powers $\psi^n$, $n \in \{1,2,3,4,5\}$, 
is reported in Figures~\ref{fig_3} and~\ref{fig:logL2_psi_n}, 
confirming rapid stabilization despite the initial memory term $w_0(x)$.

\section{Conclusions}
\label{sec:conclusion}

This paper studied a class of perturbed infinite-dimensional systems under linear time-varying feedback control. 
We established the well-posedness of the closed-loop system and demonstrated that the proposed control guarantees rapid stabilization and input-to-state stability. 
The results are further illustrated through numerical simulations, highlighting the effectiveness of the approach in achieving hyperexponential convergence despite initial perturbations.

Further perspectives include relaxing the well-posedness assumptions, extending the framework to nonlinear operators, and incorporating time-varying systems with hyper-exponential stability together with backstepping techniques for boundary control.
\bibliography{biblios}           

\end{document}